\documentclass[letterpaper]{sig-alternate-2013}

\usepackage[
  pass,  ]{geometry}

\usepackage{booktabs}

\usepackage{url}
\usepackage{graphicx}
\usepackage{multirow}
\usepackage{booktabs}
\usepackage{tabularx}
\usepackage{color, colortbl}
\usepackage{balance}
\usepackage{comment}
\usepackage{soul}
\usepackage[tight,footnotesize]{subfigure}

\definecolor{Gray}{gray}{0.85}
\usepackage{amsmath}

\def\sharedaffiliation{	\end{tabular}
	\begin{tabular}{c}
}

\permission{\copyright\ 2017 International World Wide Web Conference Committee (IW3C2), 
published under Creative Commons CC BY 4.0 License.}
\conferenceinfo{WWW 2017 Companion,}{April 3--7, 2017, Perth, Australia.}
\copyrightetc{ACM \the\acmcopyr}
\crdata{978-1-4503-4914-7/17/04. \\
http://dx.doi.org/10.1145/3041021.3055135 \\
\includegraphics{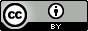}
}
 
\begin{document}
\title{The Paradigm-Shift of Social Spambots:\\Evidence, Theories, and Tools for the Arms Race}

\numberofauthors{5}
     \author{
     \alignauthor Stefano Cresci~\footnotemark[2]\enspace\footnotemark[3]\\      
     \email{s.cresci@iit.cnr.it}
     \alignauthor Roberto Di Pietro~\footnotemark[4]\enspace\footnotemark[5]\enspace\footnotemark[2]\\    
     \email{roberto.di\_pietro@nokia-bell-labs.com}	
     \alignauthor Marinella Petrocchi~\footnotemark[2]\\    
     \email{m.petrocchi@iit.cnr.it}
\and
     \alignauthor Angelo Spognardi~\footnotemark[6]\\      
     \email{angsp@dtu.dk}
     \alignauthor Maurizio Tesconi~\footnotemark[2]\\    
     \email{m.tesconi@iit.cnr.it}
     \sharedaffiliation
     \affaddr{\footnotemark[2]\quad Institute for Informatics and Telematics, National Research Council, Italy}\\
     \affaddr{\footnotemark[3]\quad Department of Information Engineering, University of Pisa, Italy}\\
     \affaddr{\footnotemark[4]\quad Nokia Bell Labs, France}\\
     \affaddr{\footnotemark[5]\quad Department of Mathematics, University of Padua, Italy}\\
     \affaddr{\footnotemark[6]\quad DTU Compute, Denmark}
}

\maketitle
\begin{abstract}
Recent studies in social media spam and automation provide anecdotal argumentation of 
the rise of a new generation of spambots, so-called \textit{social spambots}.
Here, for the first time, we extensively study this novel phenomenon on Twitter and we provide quantitative evidence that a paradigm-shift exists in spambot design. 
First, we measure current Twitter's capabilities of detecting the new social spambots. Later, we assess the human performance in discriminating between genuine accounts, social spambots, and traditional spambots.
Then, we benchmark several
state-of-the-art techniques proposed by the academic literature.
Results show that neither Twitter, nor humans, nor cutting-edge applications
are currently capable of accurately detecting the new social spambots. Our results call for new approaches capable of turning the tide in the fight against this raising phenomenon.
We conclude by reviewing the latest literature on spambots detection and we highlight an emerging common research trend based on the analysis of collective behaviors.
Insights derived from both our extensive experimental campaign and survey shed light on the most promising directions of research and lay the foundations for the arms race against the novel social spambots.
Finally, to foster research on this novel phenomenon, we make publicly available to the scientific community all the datasets used in this study.
\end{abstract}

\begin{comment}
\begin{CCSXML}
<ccs2012>
<concept>
<concept_id>10010147.10010257</concept_id>
<concept_desc>Computing methodologies~Machine learning</concept_desc>
<concept_significance>500</concept_significance>
</concept>
<concept>
<concept_id>10002978.10003022.10003027</concept_id>
<concept_desc>Security and privacy~Social network security and privacy</concept_desc>
<concept_significance>300</concept_significance>
</concept>
<concept>
<concept_id>10003120.10003130</concept_id>
<concept_desc>Human-centered computing~Collaborative and social computing</concept_desc>
<concept_significance>300</concept_significance>
</concept>
</ccs2012>  
\end{CCSXML}

\ccsdesc[500]{Computing methodologies~Machine learning}
\ccsdesc[300]{Security and privacy~Social network security and privacy}
\ccsdesc[300]{Human-centered computing~Collaborative and social computing}
\end{comment}

\vfill

\makeatletter{}
\section{Introduction}
\label{sec:introduction}

\makeatletter{}
\begin{table*}[ht]
	\scriptsize
	\centering
	\begin{tabular}{lp{0.45\textwidth}rrcc}
		\toprule
		&& \multicolumn{3}{c}{{\textbf{statistics}}} &\\
		\cmidrule{3-5}
		&&&&& \textbf{used in} \\
		\textbf{dataset} & \textbf{description} & accounts & tweets & year & \textbf{section} \\
		\midrule
		\texttt{genuine accounts}		& verified accounts that are human-operated										& 3,474	& 8,377,522	& 2011	& \ref{subsec:Twitter},~\ref{subsec:crowdsourcing}  \\
		\texttt{social spambots \#1}	& retweeters of an Italian political candidate										& 991	& 1,610,176	& 2012	& \ref{subsec:Twitter},~\ref{subsec:crowdsourcing} \\
		\texttt{social spambots \#2	}	& spammers of paid apps for mobile devices										& 3,457	& 428,542		& 2014	& \ref{subsec:Twitter},~\ref{subsec:crowdsourcing} \\
		\texttt{social spambots \#3	}	& spammers of products on sale at \textit{Amazon.com}								& 464	& 1,418,626	& 2011	& \ref{subsec:Twitter},~\ref{subsec:crowdsourcing} \\
		\texttt{traditional spambots \#1}	& training set of spammers used by Yang \textit{et al.} in~\cite{yang2013}					& 1,000	& 145,094		& 2009	& \ref{subsec:Twitter} \\
		\texttt{traditional spambots \#2}	& spammers of scam URLs													& 100	& 74,957		& 2014	& \ref{subsec:Twitter} \\
		\texttt{traditional spambots \#3}	& automated accounts spamming job offers										& 433	& 5,794,931	& 2013	& \ref{subsec:crowdsourcing} \\
		\texttt{traditional spambots \#4}	& another group of automated accounts spamming job offers							& 1,128	& 133,311		& 2009	& \ref{subsec:crowdsourcing} \\
		\texttt{fake followers}			& simple accounts that inflate the number of followers of another account				& 3,351	& 196,027		& 2012	& \ref{subsec:Twitter} \\
		\midrule
		\texttt{test set \#1}			& mixed set of 50\% \texttt{genuine accounts} + 50\% \texttt{social spambots \#1}	& 1,982	& 4,061,598	& --	& \ref{sec:theothers},~\ref{sec:newtrends} \\
		\texttt{test set \#2}			& mixed set of 50\% \texttt{genuine accounts} + 50\% \texttt{social spambots \#3}	& 928	& 2,628,181	& --	& \ref{sec:theothers},~\ref{sec:newtrends} \\
		\bottomrule
	\end{tabular}
	\caption{\small Statistics about the datasets used for this study.
	\label{tab:datasets}}
\end{table*}

The widespread availability and ease of use of Online Social Networks (OSN) have made them the ideal setting for the proliferation of fictitious and malicious accounts~\cite{liu2014}. Indeed, recent work uncovered the existence of large numbers of OSN accounts that are purposely created to distribute unsolicited spam, advertise events and products of doubtful legality, sponsor public characters and, ultimately, lead to a bias within the public opinion~\cite{ferrara2016,Jiang2016b}.
Moreover, the plague of such spammers and bots leads to an ingenious and lucrative ``underground economy", where account vendors, their customers, and oblivious victims play a piece staging since the very introduction of social networks~\cite{Stringhini:2012,Stringhini:2013,Thomas2013}. 

One of the most fascinating peculiarities of spambots is that they ``evolve" over time, adopting sophisticated techniques to evade early-established detection approaches, such as those based on textual content of shared messages~\cite{Lee:2010}, posting patterns~\cite{Stringhini:2010} and social relationships~\cite{Ghosh:2012}. As evolving spammers became clever in escaping detection, for instance by changing discussion topics and posting activities, researchers kept the pace and proposed complex models, such as those based on the interaction graphs of the accounts under investigation~\cite{yang2013, hu2014}.

Noticeably, spambots evolution still goes on. Recent investigations anecdotally highlight how new waves of \textit{social spambots} are rising~\cite{ferrara2016, zhang2016}. In this paper, we target these new waves, finding evidence of the difficulties for OSN users to distinguish between genuine and malicious accounts. We also highlight the difficulties for OSN administrators to take appropriate countermeasures against the takeover of evolving spambots.
Remarkably, a large number of tools and techniques have been proposed by Academia to detect OSN spambots~\cite{ferrara2016,Jiang2016b}. Until recently, such tools have proved to be valid allies for spambots timely detection. 
Unfortunately, the characteristics of the new wave of social spambots are such that standard classification approaches, where a single account is evaluated according to a set of established features tested over known datasets, are no longer successful. In this work, we demonstrate this claim by investigating the performances of several state-of-the-art tools techniques when struggling against the latest wave of social spambots.
The unsatisfactory results of the surveyed techniques call for new approaches capable of turning the tide of this long-lasting fight.

Interestingly, we assist to a paradigm-shift in modeling
and analyzing online accounts.
Independently from each other, new research efforts were born, which leverage characteristics of groups of accounts -- rather than those of a single account -- as a red flag for anomalous behaviors. We provide a review of these prominent research directions, highlighting the new dimensions to sound out for successfully fighting against  this novel generation of spambots.

\makeatletter{}
\begin{table*}[t]
	\scriptsize
	\centering
	\begin{tabular}{lrrrr}
		\toprule
		& \multicolumn{4}{c}{{\textbf{accounts}}}\\
		\cmidrule{2-5}
		\textbf{dataset} & total & alive & deleted & suspended \\
		\midrule
		\texttt{genuine accounts}		& 3,474	& 3,353 (96.5\%)	& 115 (3.3\%)	& 6 (0.1\%) \\
		\texttt{social spambots \#1}	& 994	& 946 (95.2\%)		& 2 (0.2\%)	& 46 (4.6\%) \\
		\texttt{social spambots \#2	}	& 3,457	& 3,322 (96.1\%)	& 1 (0.1\%)	& 134 (3.8\%) \\
		\texttt{social spambots \#3	}	& 467	& 465 (99.6\%)		& 2 (0.4\%)	& 0 (0.0\%) \\
		\texttt{traditional spambots \#1}	& 1,000	& 889 (88.9\%)		& 25 (2.5\%)	& 86 (8.6\%) \\
		\texttt{traditional spambots \#2}	& 100	& 1 (1.0\%)		& 0 (0.0\%)	& 99 (99.0\%) \\
		\texttt{fake followers}			& 3,351	& 851 (25.4\%)		& 38	(1.1\%)	& 2,462 (73.5\%) \\
		\bottomrule
	\end{tabular}
	\caption{\small Statistics about alive, deleted, and suspended accounts, for different groups of genuine and malicious accounts.
	\label{tab:survivability}}
\end{table*}

\vskip 1em
\noindent \textbf{Contributions.} Our main contributions are:
\begin{itemize}
\item We provide empirical evidence of the existence of a novel wave of Twitter spambots, which, up to now, has been just theorized~\cite{ferrara2016}. \item We evaluate if, and to which extent, state-of-the-art detection techniques succeed in spotting such new spambots. \item We critically revise an emerging stream of research, which adopt features tied to groups of accounts rather than individual accounts features. \item We leverage results of a crowdsourcing spambot detection campaign for drawing new guidelines for the annotation of datasets comprising social spambots.
\item Finally, we publicly release to the scientific community an annotated dataset\footnote{\scriptsize{Available at: \url{http://mib.projects.iit.cnr.it/dataset.html}}}, consisting of genuine accounts, traditional spambots, and -- for the first time -- the novel social spambots.  \end{itemize}

%%% End:  

\makeatletter{}
\section{Datasets}
\label{sec:datasets}
We describe the different Twitter
datasets that constitute the real-world data used in our
experiments.
Table~\ref{tab:datasets} reports the name of the datasets, their brief description, and the number of accounts and tweets they feature. The year represents the average of the creation years of the accounts that belong to the dataset. 

The \texttt{genuine accounts} dataset is a random sample of genuine (human-operated) accounts. Following a hybrid crowdsensing approach~\cite{avvenuti2017}, we randomly contacted Twitter users by 
asking them a simple question in natural language. All the replies to our questions were manually verified and all the 3,474 accounts that answered were certified as humans. The accounts that did not answer to our question were discarded and are not used in this study.

The \texttt{social spambots \#1} dataset was created after observing the activities of a novel group of social bots that we discovered on Twitter during the last Mayoral election in Rome, in 2014. One of the runners-up employed a social media marketing firm for his electoral campaign, which made use of almost 1,000 automated accounts on Twitter to publicize his policies. Surprisingly, we found such automated accounts to be similar to genuine ones in every way. Every profile was accurately filled in with detailed -- yet fake -- personal information such as a (stolen) photo, (fake) short-bio, (fake) location, etc. Those accounts also represented credible sources of information since they all had thousands of followers and friends, the majority of which were genuine users\footnote{\scriptsize{This was made possible also by the adoption of social engineering techniques, such as the photo of a young attractive woman as the profile picture and the occasional posting of provocative tweets.}}. Furthermore, the accounts showed a tweeting behavior that was apparently similar to those of genuine accounts, with a few tweets posted every day, mainly quotes from popular people, songs, and \textit{YouTube} videos. However, every time the political candidate posted a new tweet from his official account, all the automated accounts retweeted it in a time span of just a few minutes. 
Thus, the political candidate was able to reach many more accounts in addition to his direct followers and managed to alter Twitter engagement metrics during the electoral campaign.
Amazingly, we also found tens of human accounts who tried to engage in conversation with some of the spambots. The most common form of such human-to-spambot interaction was represented by a human reply to one of the spambot tweets quotes. 
We also discovered a second group of social bots, which we labeled \texttt{social spambots \#2}, who spent several months promoting the \verb|#TALNTS| hashtag. Specifically, \textit{Talnts} is a mobile phone application for getting in touch with and hiring artists working in the fields of writing, digital photography, music, and more. The vast majority of tweets 
were harmless messages, occasionally interspersed by tweets mentioning a specific genuine (human) account and suggesting him to buy the VIP version of the app from a Web store. 

Further, we uncovered a third group of social bots, \texttt{social spambots \#3}, which advertise products on sale on \textit{Amazon.com}. The deceitful activity was carried out by spamming URLs pointing to the advertised products. Similarly to the retweeters of the Italian political candidate, also this family of spambots interleaved spam tweets with harmless and genuine ones.

We exploited a Twitter crawler to
collect data about all the accounts we suspected to belong to the three
groups of social spambots. All the accounts collected in this process have
then undergone an internal manual verification phase to certify their automated
nature. Among all the distinct retweeters of the Italian political
candidate, 50.05\% (991 accounts) were certified as
spambots. Similarly, 94.50\% (3,457 accounts) of the accounts who tweeted the \verb|#TALNTS| hashtag resulted as spambots. Finally, 89.29\% (464 accounts) of the accounts that
tweeted suspicious \textit{Amazon.com} URLs were also certified as
spambots. The three sets of accounts represent our ground truth of novel social spambots. 

Our internal manual annotation has been carried out by comparing every account to all the others, in order to highlight possible similarities and common behaviors. This is in contrast with the typical annotation process where accounts are labeled one-by-one and by solely exploiting the characteristics of the account under investigation.

In addition to genuine users and social spambots, we also collected several datasets of traditional spambots. Such datasets are used throughout the paper as a strong baseline. The \texttt{traditional spambots \#1} dataset is the training set used in~\cite{yang2013}, kindly provided to us by the authors of that work. In~\cite{yang2013}, the dataset has been used to train a machine learning classifier for the detection of evolving Twitter spambots.
Accounts belonging to the \texttt{traditional spambots \#2} dataset are rather simplistic bots that repeatedly mention other users in tweets containing scam URLs. To lure users into clicking the malicious links, the content of their tweets invite the mentioned users to claim a monetary prize.
The \texttt{traditional spambots \#3} and \texttt{traditional spambots \#4} datasets are related to 2 different groups of bots that repeatedly tweet about open job positions and job offers.

Fake followers are another kind of malicious accounts that recently gained interest both from platform administrators and from the scientific world~\cite{cresci2015}. Given that fake followers are rather simplistic in their design and functioning, they can serve as a weak baseline against which to compare social spambots.
In April, 2013, we bought 3,351 fake accounts from three different Twitter online markets, namely \textit{fastfollowerz.com}, \textit{intertwitter.com}, and \textit{twittertechnology.com}. All the accounts acquired in this way have been merged in order to obtain the \texttt{fake followers} dataset used in this study.

By considering a diverse set of spammer accounts we have captured many of the different dimensions currently exploited by spambots and tamperers to perpetrate their illicit activities. In detail, we have considered (i) fake follower frauds, (ii) retweet frauds, (iii) hashtag promotion, (iv) URL spamming, (v) scamming, and (vi) spam of generic messages.

\begin{comment}
{\bf mari: x quali accounts vale questa parte qua sotto?}
For all the
4,929 accounts of our datasets we then collected behavioral data by
crawling the content of their Twitter pages. Furthermore, we also
collected data about all their direct followers and friends, and about
all the accounts they interacted with in their
tweets. 
\end{comment}

%%% End:  

\makeatletter{}
\section{Real-world experimentation}
\label{sec:realworld}

\subsection{Twitter monitoring}
\label{subsec:Twitter}

A first assessment of the extent and the severity of Twitter social spambots problem can be obtained by measuring Twitter's capacity of detecting and removing them from the platform. This section thus answers the research question:

\textbf{RQ1 --} \textit{To what extent is Twitter currently capable of detecting and removing social spambots?}

Interesting insights can be gained by comparing the rate at which Twitter accounts are removed, for different types of malicious accounts. The intuition is that accounts that are easily identified as malicious can be rapidly removed by platform administrators. Thus, in this experiment, we let different types of accounts behave for a rather long amount of time (i.e., years). Then, we check whether Twitter managed to identify such accounts as malicious and to remove them from the platform. We perform this experiment on our set of genuine accounts, on our 3 groups of social spambots, on 2 groups of traditional spambots, and on the group of fake followers.

\begin{figure}
  \centering
{\includegraphics[width=0.995\columnwidth]{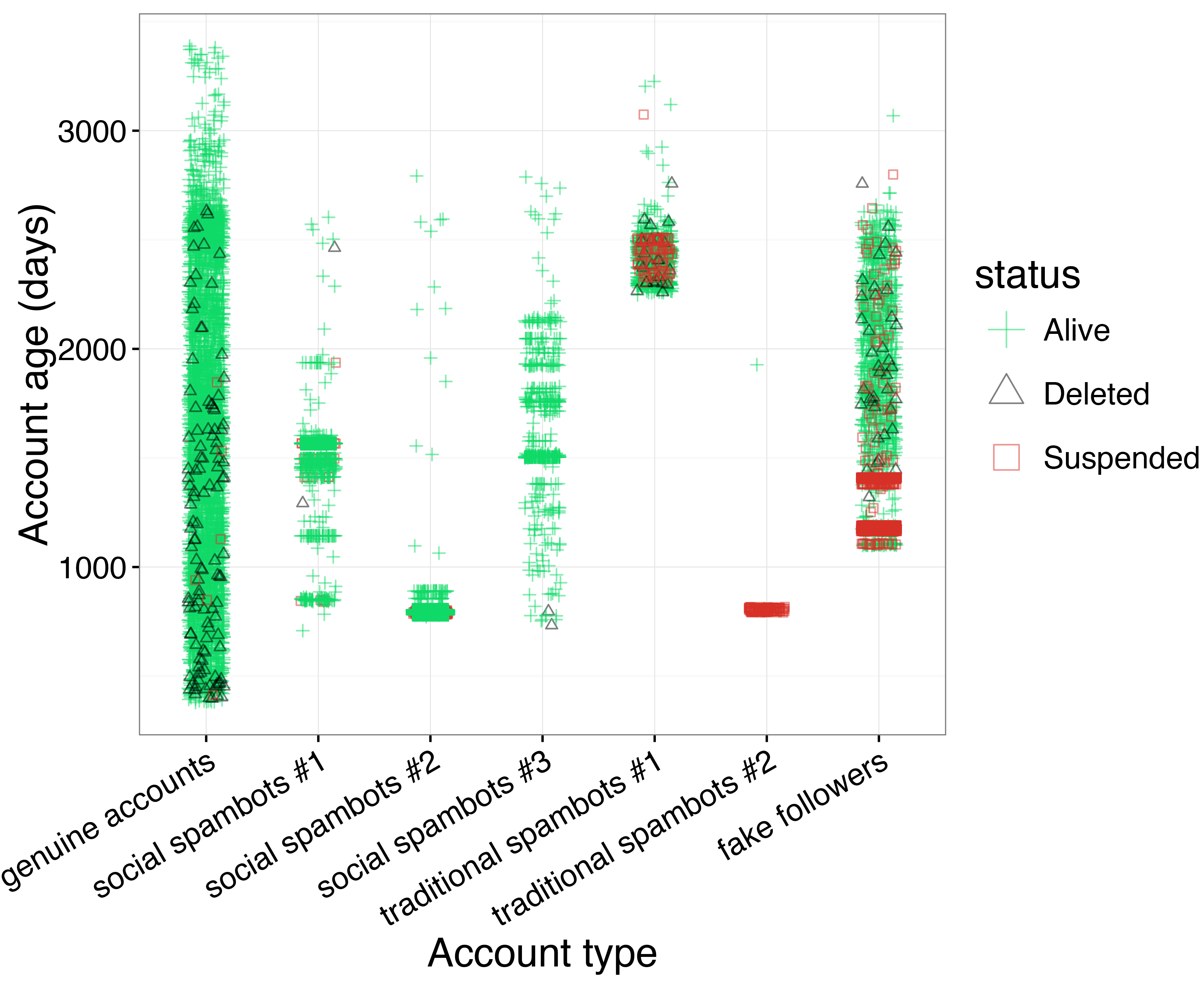}}
 \caption{\small Survival rates for different types of accounts.\label{fig:survivability}}
\end{figure}

In order to perform this experiment, we exploited Twitter's responses to API calls and, particularly, the Twitter error codes.
Given a query to a specific account, Twitter's API replies with information regarding the status of the queried account. Specifically, accounts that are suspected to perform malicious activities get suspended by Twitter. API queries to a suspended account result in Twitter responding with the error code 63. API queries to accounts that have been deleted by their original owner result in Twitter responding with the error code 50. Instead, for accounts that are neither suspended nor deleted, Twitter replies with the full metadata information of the account, without issuing error codes. By exploiting this response mechanism, we were able to measure the \textit{survivability} of the different groups of accounts. Results of this experiment are reported in Table~\ref{tab:survivability} and are pictorially depicted in Figure~\ref{fig:survivability}.

As shown in Table~\ref{tab:survivability}, \texttt{genuine accounts} feature a very high survival rate (96.5\%). In addition, among the no longer available accounts, the vast majority have been deleted by the original owner, rather than suspended by Twitter. These results are quite intuitive, by considering that legitimate accounts rarely perform any kind of malicious activity. Conversely, the simplest kind of malicious accounts, \texttt{fake followers}, have mostly been detected and suspended by Twitter. The same also applies to one of the two groups of traditional spambots, identified as \texttt{traditional spambots \#2} in Table~\ref{tab:survivability}, which features a suspension rate as high as 99\%. The most interesting results are however related to those kinds of malicious accounts that better mimic human behaviors. So far, \texttt{traditional spambots \#1} have largely managed to evade suspension, despite dating back to 2009. Indeed, only 8.6\% of the bots have been suspended, while 88.9\% of them are still alive. This seems to suggest that Twitter's spambot detection mechanisms are still unable to accurately identify such accounts, while recent solutions proposed by Academia have succeeded in this task~\cite{yang2013}. Twitter's performance in suspending malicious accounts is even worse if we consider social spambots. All the 3 groups of social spambots feature very high survival rates, respectively 95.2\%, 96.1\%, and 99.6\%. Even if the difference between the survival rate of social spambots and that of \texttt{traditional spambots \#1} is marginal, these results nonetheless suggest an increased difficulty for the detection of social spambots. Table~\ref{tab:surv-stat-sign} also reports the results of a comparison between the ratios of alive, deleted, and suspended accounts between spambots and \texttt{genuine accounts}. As shown, social spambots feature very small differences with respect to genuine accounts ($\sim\pm3\%$). Some of these differences are not even statistically significant, according to a chi-square test. \texttt{Traditional spambots \#1} have differences $\sim\pm8\%$ that are highly significant ($p < 0.01$) for alive and suspended accounts.
Instead, \texttt{traditional spambots \#2} and \texttt{fake followers} show massive differences: $\sim\pm96\%$ and $\sim\pm72\%$, respectively.

Figure~\ref{fig:survivability} shows results of the survivability experiment, with respect to the account age\footnote{\scriptsize{Account age is computed as the number of days between the account's creation date and the day we performed the experiment.}}.
This can allow to understand if temporal patterns exist in the way malicious accounts are created, and if Twitter's mechanisms for suspending malicious accounts are related to an account's age. For instance, Twitter might be better in detecting and suspending older accounts than newer ones. However, this hypothesis can be ruled out by considering that 99\% of \texttt{traditional spambots \#2} accounts have been suspended despite being younger than most of the social spambots.
Overall, an analysis of Figure~\ref{fig:survivability} shows that account suspensions seem to depend on the type of the account, its design and behavior, rather than on its age.

\makeatletter{}
\begin{table}[t]
	\scriptsize
	\centering
	\begin{tabular}{lr@{}l r@{}l r@{}l}
		\toprule
		& \multicolumn{6}{c}{{\textbf{accounts}}}\\
		\cmidrule{2-7}
		\textbf{dataset} & \multicolumn{2}{c}{alive} & \multicolumn{2}{c}{deleted} & \multicolumn{2}{c}{suspended} \\
		\midrule
		\texttt{social spambots \#1}	& $-$1.&3\%\textsuperscript{*}		& $-$3.&1\%\textsuperscript{***}	& $+$4.&5\%\textsuperscript{***} \\
		\texttt{social spambots \#2	}	& $-$0.&4\%					& $-$3.&2\%\textsuperscript{***}	& $+$3.&7\%\textsuperscript{***} \\
		\texttt{social spambots \#3	}	& $+$3.&1\%\textsuperscript{***}	& $-$2.&9\%\textsuperscript{***}	& $-$0.&1\% \\
		\texttt{traditional spambots \#1}	& $-$7.&6\%\textsuperscript{***}	& $-$0.&8\%					& $+$8.&7\%\textsuperscript{***} \\
		\texttt{traditional spambots \#2}	& $-$95.&5\%\textsuperscript{***}	& $-$3.&3\%					& $+$98.&9\%\textsuperscript{***} \\
		\texttt{fake followers}			& $-$71.&1\%\textsuperscript{***}	& $-$2.&2\%\textsuperscript{***}	& $+$73.&4\%\textsuperscript{***} \\
		\bottomrule
		\multicolumn{4}{l}{\rule{0pt}{1.2\normalbaselineskip}
		{\scriptsize \textsuperscript{***}$p < 0.01$, \textsuperscript{**}$p < 0.05$, \textsuperscript{*}$p < 0.1$}}
	\end{tabular}
	\caption{\small Effect size and statistical significance of the difference between the survivability results of malicious accounts with respect to those of \texttt{genuine accounts}.
	\label{tab:surv-stat-sign}}
\end{table}

Results reported in this first experiment  already reveal interesting differences between social spambots, traditional spambots, and fake followers.
Notably, social spambots appear to be more similar to genuine accounts than to traditional spambots, with regards to Twitter suspensions.
\subsection{Crowdsourcing: tasks and results}
\label{subsec:crowdsourcing}
This section addresses the following research questions: 

\textbf{RQ2 --} \textit{Do humans succeed in detecting social spambots in the wild?}

\textbf{RQ3 --} \textit{Do they succeed in discriminating between traditional spambots, social spambots, and genuine accounts?}

Even if Twitter users were generally capable of distinguishing between traditional spambots and genuine accounts, they might still find it difficult to spot social spambots in the wild. If confirmed, this would provide additional evidence of the evolutionary step characterizing the new social spambots with respect to traditional ones.
 
 \begin{figure}[h]
  \centering
{\includegraphics[width=0.995\columnwidth]{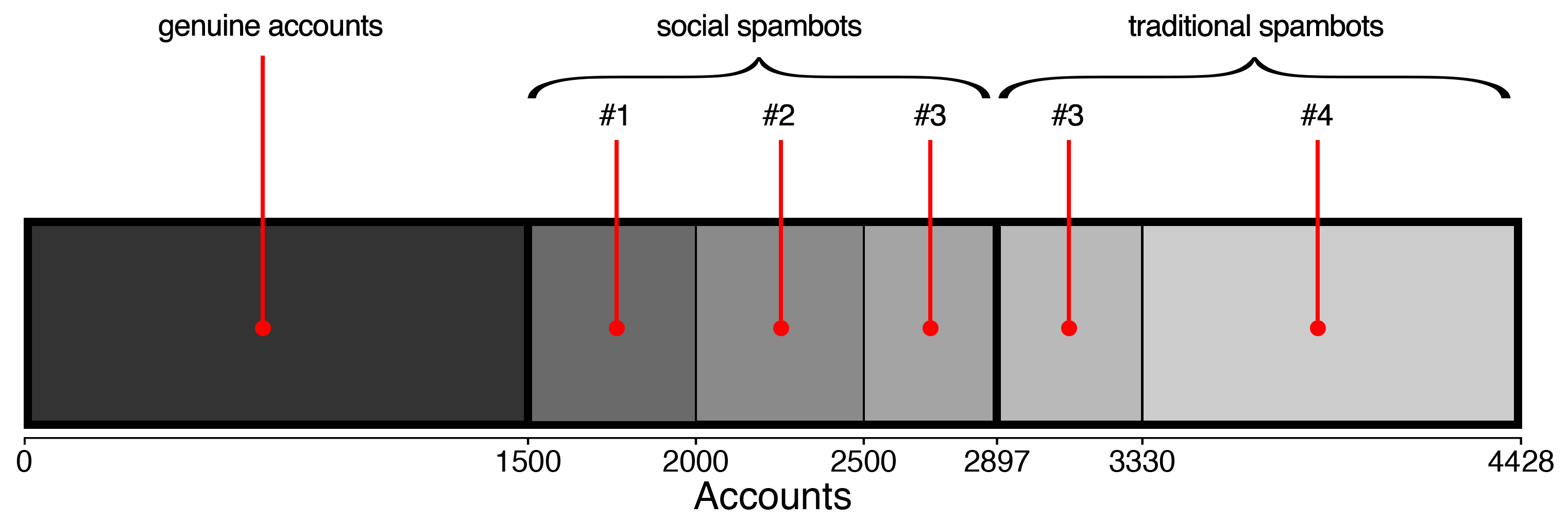}}
 \caption{\small Dataset composition for the crowdsourcing experiment.\label{fig:crowdflower}}
\end{figure}
 
To answer these research questions, we asked a large set of real-world users to classify the accounts in our datasets.
To obtain a large and diverse set of users, we recruited contributors from the CrowdFlower\footnote{\scriptsize{\url{https://www.crowdflower.com/}}} crowdsourcing platform.
Figure~\ref{fig:crowdflower} shows the distribution of the 4,428 accounts that we have employed for this crowdsourcing experiment, picked up from the datasets in Section~\ref{sec:datasets}.
Contributors were asked to assign to each account one of the following classes: (i) spambot, (ii) genuine, and (iii) unable to classify. The latter class (iii) has been inserted to deal with Twitter accounts possibly getting deleted, suspended, or protected\footnote{\scriptsize{\textit{Protected} accounts are those accounts whose tweets and timeline are not publicly visible.}} while our crowdsourcing task was ongoing.

Notably, our experiment marks a difference with those typically carried out with crowdsourcing. In fact, crowdsourcing tasks are typically aimed at creating a ground truth (i.e., labeled) dataset for later use. For instance, crowdsourcing is often used to create large training-sets for machine learning algorithms. Here, instead, the datasets are labeled in advance. Thus, by asking contributors to (re-)classify our datasets, we are actually evaluating their ability to spot the different types of accounts.

\makeatletter{}
\begin{table}[t]
	\scriptsize
	\centering
	\begin{tabular}{lr}
		\toprule
		Num. accounts to classify		& 4,428 \\%4,453\\
		Min. contributors per account	& 3 \\
		Max. answers per contributor	& 100 \\
		Num. test questions         		& 25    \\
		Min. accuracy threshold		& 70\% \\
		Reward					& 0.1 US\$ per 5 accounts classified \\

		\bottomrule
	\end{tabular}
\caption{\small Crowdsourcing campaign settings.
\label{tab:crowdsetting}}
\end{table}

%%% End:  

\vskip 1em
\noindent \textbf{Enforcing results reliability.} We only recruited contributors who were tech-savvy and Twitter users themselves, in order to be reasonably sure about their knowledge of Twitter and its dynamics.
Furthermore, we required each account to be classified by at least 3 different contributors, with the final class decided by majority voting.
We also fixed to 100 the upper threshold of the number of accounts that a single contributor could classify. In this way, we have obtained redundant results from a broad set of contributors. Then, in order to further guarantee the reliability of our crowdsourcing results, we designed a set of ``test" (or ``gold") questions aimed at evaluating the quality of contributors' answers. A test question is one for which the correct answer is already known by the system. Within the crowdsourcing platform, such questions are indistinguishable from standard ones and are randomly mixed among all the questions, so that contributors cannot know whether they are answering to a test or to a standard question. Contributors' answers to test questions were checked against the known correct answers. Only the trusted contributors who answered correctly to more than the 70\% of the test questions
have been considered in our study. Our test questions consist of accounts whose nature is ``easily" recognizable, and specifically: (i) a set of traditional spambots sampled from the dataset of Yang {\it et al.}~\cite{yang2013}, (ii) a subset of genuine accounts, and (iii) a set of suspended, deleted, and protected accounts.
Notably, by designing test questions with traditional spambots and genuine accounts, and by enforcing the policy of at least 70\% correct answers, we can guarantee that all our trusted contributors are typically able to detect traditional spambots and to distinguish them from genuine accounts. This further strengthens the results of their classification of the novel social spambots.

\begin{figure}[t]
\centering
\subfigure[t][Answers per country (top 20).\label{fig:crowdcountries}]{\includegraphics[width=0.65\columnwidth]{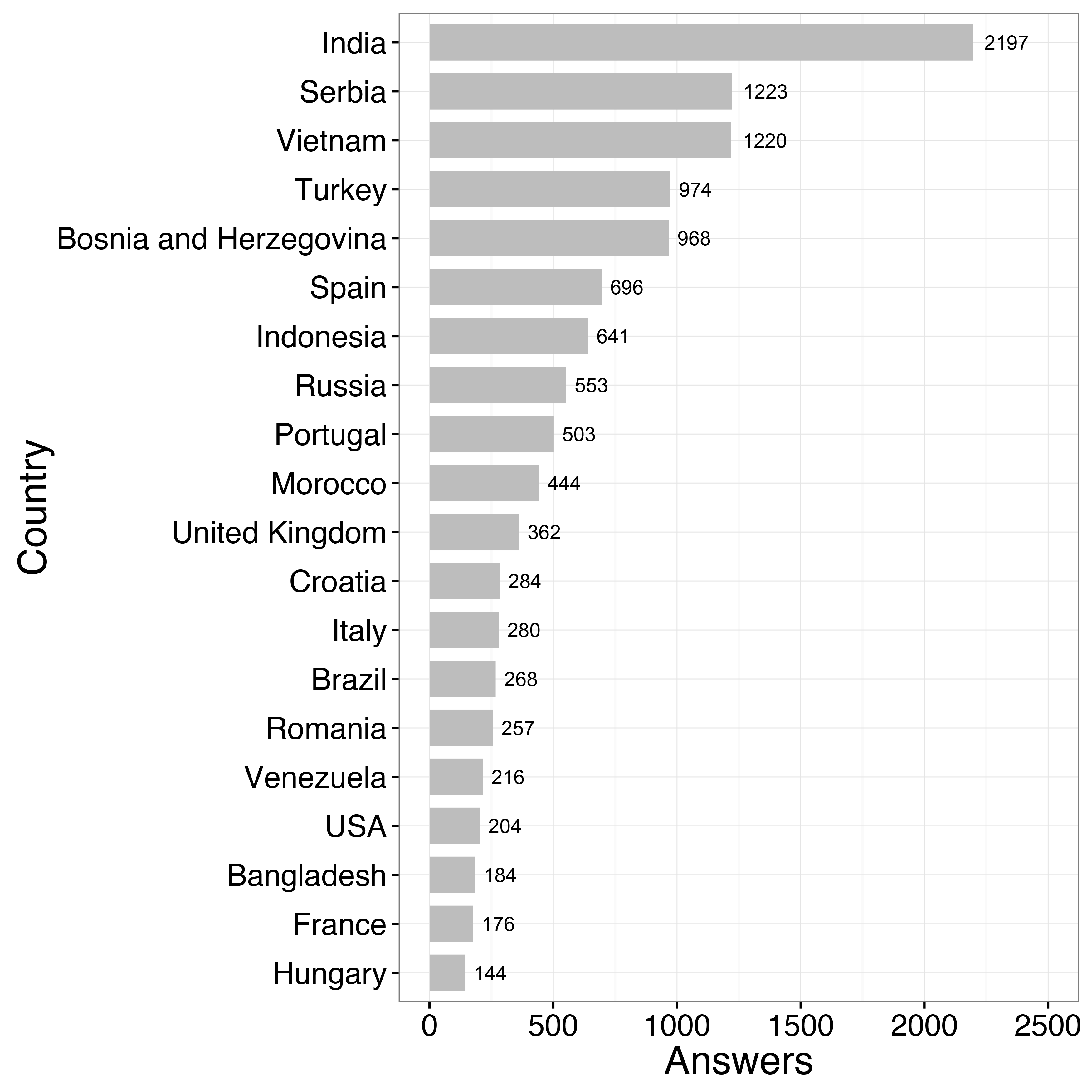}}\\
\subfigure[t][Answers per contributor.\label{fig:crowdcontrib}]{\includegraphics[width=0.75\columnwidth]{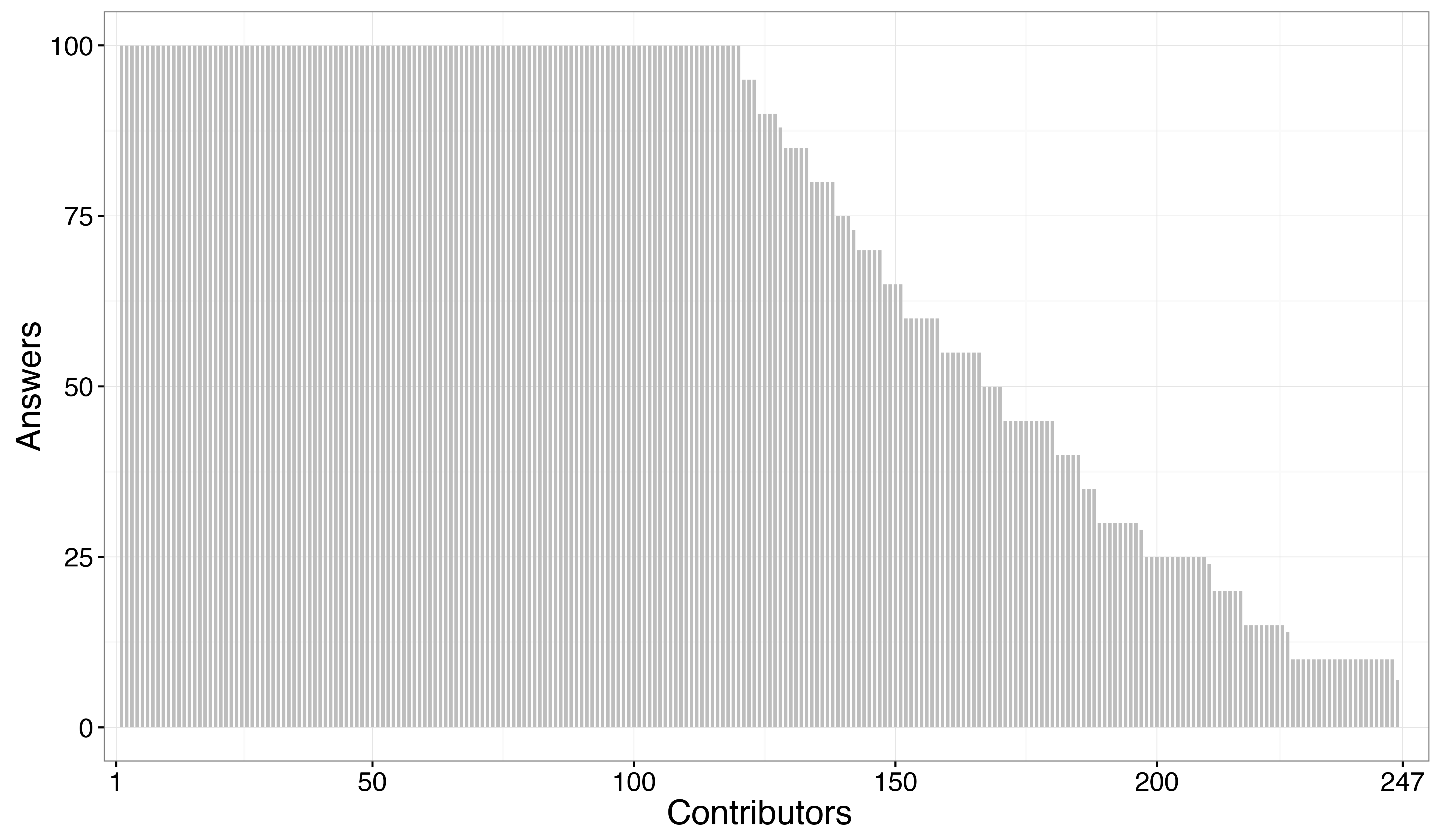}}
\caption{\label{fig:crowd}Distribution of crowdsourcing results.}
\end{figure}
\makeatletter{}
\begin{table}[t]
	\scriptsize
	\centering
	\begin{tabular}{lr}
		\toprule
		Instructions clear		& 4.0 / 5 \\
		Test questions fair		& 3.5 / 5 \\
		Ease of job			& 3.5 / 5 \\
		Pay		         		& 3.8 / 5 \\
		\midrule
		Overall				& 3.7 / 5 \\
		\bottomrule
	\end{tabular}
\caption{\small Contributors' evaluation of our campaign.
\label{tab:crowdscore}}
\end{table}

%%% End:  

Table~\ref{tab:crowdsetting} shows a recap of the settings used in our crowdsourcing campaign.
The thorough description of our campaign, with the complete set of instructions, a list of example accounts, and the task preview, is available online\footnote{\scriptsize{\url{http://wafi.iit.cnr.it/fake/fake/crowdflower/instructions/}}}.
The campaign completed when each of the 4,428 accounts was classified by 3 different trusted contributors.

\makeatletter{}
\begin{table*}[t]
	\scriptsize
	\centering
	\begin{tabular}{lrrrrrrrrrr}
		\toprule
		&&& \multicolumn{5}{c}{{\textbf{detection results}}} &&& \\
		\cmidrule{4-8}
		\textbf{type} & \textbf{accounts $^{\sharp}$} && TP & TN & FP & FN & Accuracy && \textbf{Fleiss' kappa ($\kappa$)} \\
		\midrule
		traditional spambots			& 1,516	&& 1,385	& 0		& 0		& 131	& 0.9136	&& 0.007 \\
		social spambots			& 1,393	&& 328	& 0		& 0		& 1,065	& 0.2355	&& 0.186 \\
		\texttt{genuine accounts}		& 1,377	&& 0		& 1,267	& 110	& 0		& 0.9201	&& 0.410 \\
		\bottomrule
		\multicolumn{11}{l}{
		\begin{minipage}[t]{1.5\columnwidth}			{\scriptsize $^{\sharp}$: The total number of accounts considered is 4,286 instead of 4,428 because 142 accounts (3.2\%) got deleted, suspended, or protected during our campaign.} 		\end{minipage}}
	\end{tabular}
	\caption{\small Results of the crowdsourcing campaign on spambots detection.
	\label{tab:crowdresults}}
\end{table*}

\vskip 1em
\noindent \textbf{Results of the crowdsourcing campaign.} Overall, we collected 13,284 answers given by 247 trusted contributors from 42 different countries.
Figure~\ref{fig:crowdcountries} shows the distribution of answers per country, while Figure~\ref{fig:crowdcontrib} depicts the distribution of answers per contributor.
CrowdFlower also gives contributors the possibility to evaluate crowdsourcing campaigns for: (i) clarity of instructions, (ii) fairness of the test questions, (iii) ease of the task, and (iv) appropriateness of the payment. Out of the 247 participating contributors, 60 of them ($\sim24\%$) evaluated our campaign, leading to a convincing aggregated score of 3.7/5, as shown in detail in Table~\ref{tab:crowdscore}.
Our campaign costed us 410 US\$ in total.

\begin{comment}
\begin{figure}[t]
  \centering
{\includegraphics[width=0.30\textwidth]{img/answers_per_country_top20.png}}
 \caption{\small Answers per country (top 20 countries).\label{fig:crowdcountries}}
\end{figure}

\begin{figure}[t]
  \centering
{\includegraphics[width=0.35\textwidth]{img/answers_per_contributor.png}}
 \caption{\small Answers per contributor.\label{fig:crowdcontrib}}
\end{figure}
\end{comment}

The most interesting results of our crowdsourcing campaign are undoubtedly related to the detection performance of our human contributors.
As reported in Table~\ref{tab:crowdresults}, overall, the human annotators obtained an accuracy of less than 0.24 on the social spambots, with more than 1,000 False Negatives (FN), meaning that contributors classified more than 1,000 accounts as genuine, when they actually belonged to the dataset of the last generation of spambots. Human detection performances for the two other groups of accounts, namely traditional spambots and genuine accounts, are instead quite satisfactory, with an accuracy of 0.91 and 0.92 respectively. These important results further highlight the existence of a striking difference between traditional and social spambots. More worryingly, they also suggest that humans might not be able to detect social spambots in the wild, and to distinguish them from genuine accounts.

Given that each account under investigation has been classified by at least 3 different contributors, we have also computed the Fleiss' kappa ($\kappa$) inter-rater agreement metric~\cite{gwet2014}. All inter-rater agreement metrics measure the level of agreement of different annotators on a task. The level of agreement can be also interpreted as a proxy for the difficulty of a task. In our experiment, human contributors showed a decent agreement for the classification of genuine accounts, with $\kappa = 0.410$. Instead, they showed very little agreement while classifying traditional spambots, as represented by $\kappa = 0.007$. This interesting result shows that, overall, the human contributors were able to correctly detect traditional spambots, as shown by the 0.91 accuracy, but also that contributors rarely agreed on the class. Surprisingly, we measured a slightly higher agreement for the classification of social spambots than for traditional ones, with $\kappa = 0.186$. These results imply that humans generally failed in classifying social spambots (accuracy $= 0.2355$) and, furthermore, that they also were more in agreement on this mistake than they were when (correctly) classifying traditional spambots.

\makeatletter{}
\begin{table*}[t]
	\scriptsize
	\centering
	\begin{tabular}{llrrrrrr}
		\toprule
		&& \multicolumn{6}{c}{\textbf{detection results}} \\
		\cmidrule{3-8}
		\textbf{technique} & \textbf{type} & Precision & Recall & Specificity & Accuracy & F-Measure & MCC \\
		\midrule
		\multicolumn{8}{l}{\texttt{test set \#1}} \\  
			Twitter countermeasures						& mixed		& \textbf{1.000}	& 0.094	& \textbf{1.000}	& 0.691	& 0.171	& 0.252 \\
			Human annotators							& manual		& 0.267	& 0.080	& 0.921	& 0.698	& 0.123	& 0.001 \\
			BotOrNot?~\cite{davis2016}					& supervised	& 0.471	& 0.208	& 0.918	& 0.734	& 0.288	& 0.174 \\
			C. Yang \textit{et al.}~\cite{yang2013}			& supervised	& 0.563	& 0.170	& 0.860	& 0.506	& 0.261	& 0.043 \\  
			Miller \textit{et al.}~\cite{miller2014}				& unsupervised	& 0.555	& 0.358	& 0.698	& 0.526	& 0.435	& 0.059 \\  
			Ahmed \textit{et al.}~\cite{ahmed2013} $^{\sharp}$	& unsupervised	& 0.945	& 0.944	& 0.945	& 0.943	& 0.944	& 0.886 \\  
						Cresci \textit{et al.}~\cite{IntSys2015}			& unsupervised	& 0.982	& \textbf{0.972}	& 0.981	& \textbf{0.976}	& \textbf{0.977}	& \textbf{0.952} \\  
		\midrule
		\multicolumn{8}{l}{\texttt{test set \#2}} \\  
			Twitter countermeasures						& mixed		& \textbf{1.000}	& 0.004	& \textbf{1.000}	& 0.502	& 0.008	& 0.046 \\
			Human annotators							& manual		& 0.647	& 0.509	& 0.921	& 0.829	& 0.570	& 0.470 \\
			BotOrNot?~\cite{davis2016}					& supervised	& 0.635	& \textbf{0.950}	& 0.981	& 0.922	& 0.761	& 0.738 \\
			C. Yang \textit{et al.}~\cite{yang2013}			& supervised	& 0.727	& 0.409	& 0.848	& 0.629	& 0.524	& 0.287 \\  
			Miller \textit{et al.}~\cite{miller2014}				& unsupervised	& 0.467	& 0.306	& 0.654	& 0.481	& 0.370	& -0.043 \\  
			Ahmed \textit{et al.}~\cite{ahmed2013} $^{\sharp}$	& unsupervised	& 0.913	& 0.935	& 0.912	& 0.923	& \textbf{0.923}	& 0.847 \\  
						Cresci \textit{et al.}~\cite{IntSys2015}			& unsupervised	& \textbf{1.000}	& 0.858	& \textbf{1.000}	& \textbf{0.929}	& \textbf{0.923}	& \textbf{0.867}  \\  
		\bottomrule
		\multicolumn{8}{l}{
		\begin{minipage}[t]{1.25\columnwidth}			{\scriptsize $^{\sharp}$: Modified by employing \textit{fastgreedy} instead of \textit{MCL} for the graph clustering step.} 		\end{minipage}}
	\end{tabular}
	\caption{\small Comparison among the spambot detection techniques, tools, and algorithms surveyed in this study. For each test set, the highest values in each evaluation metric are shown in bold.}
	\label{tab:results}
\end{table*}

\vskip 1em
\noindent \textbf{Annotation guidelines for spambots detection.} Despite the recent advances in machine learning-based detection systems, manual verification of accounts to assess their degree of automation is still carried out by platform administrators~\cite{ferrara2016}. 
  In~\cite{wang2013}, it is reported that human experts ``consistently produced near-optimal results" on a dataset of traditional spambots.
However, the results of our crowdsourcing experiment confirmed that the traditional ``account-by-account" annotation process used by human workers to evaluate social media accounts is no longer viable when applied to the detection of the novel wave of social spambots. Given the importance of manual annotation for the creation of ground-truth datasets and for double-checking suspicious accounts on social networking platforms, we call for the adoption of new annotation methodologies that take into account the similarities and synchronized behaviors of the accounts. We have adopted a practical implementation of this methodology to annotate our datasets of social spambots. In particular, we have compared the timelines of large groups of accounts, in order to highlight tweeting similarities among them. By comparing the behaviors of different accounts, rather than by analyzing them one by one, we were able to spot the social spambots among all the collected accounts, as thoroughly described in Section~\ref{sec:datasets}. Therefore, we envisage the possibility to adopt this methodology, as well as similar ones, in order to safeguard the manual annotation process from elusive social spambots.

%%% End:  

\makeatletter{}
\section{Established techniques}
\label{sec:theothers}
So far, we have demonstrated that neither Twitter nor human operators are currently capable of identifying novel social spambots.
Here, we investigate whether established tools and techniques
are able to succeed in this task. Thus, our research question is:

\textbf{RQ4 --} \textit{Are state-of-the-art scientific applications and techniques able to detect social spambots?}

\vskip 1em
\noindent \textbf{The \textit{BotOrNot?} service.} BotOrNot? is a publicly-available service\footnote{\scriptsize{\url{http://truthy.indiana.edu/botornot/}}} to evaluate the similarity of a Twitter account with the known characteristics of social spambots~\cite{davis2016}. It has been developed by the Indiana University at Bloomington and it was released in May 2014.
Claimed capable of detecting social spambots~\cite{ferrara2016}, at the time of writing it was the only publicly-available social spambot detection system. 
BotOrNot? leverages a supervised machine-learning classifier that exploits more than 1,000 features of the Twitter account under investigation. Specifically, it employs off-the-shelf supervised learning algorithms trained with examples of both humans and bots behaviors, based on the Texas A\&M dataset~\cite{leeKyumin2011} with 15,000 examples of each class and millions of tweets. 
Similarly to most already established techniques, BotOrNot? performs its analyses on an account-by-account basis. 
Despite being specifically designed for the detection of social spambots, authors state that the detection performances of BotOrNot? against evolved spambots might be worse than those reported in~\cite{davis2016}. Here, we aim at evaluating this point by querying the BotOrNot? service with our sets of genuine and social spambot accounts.
As shown in Table~\ref{tab:results}, BotOrNot? achieves rather unsatisfactory results for the accounts of both \texttt{test set \#1} and \texttt{test set \#2} (such datasets are described in Table~\ref{tab:datasets}). Its detection performances are particularly bad for the accounts of \texttt{test set \#1} -- where the spambots are from the \texttt{social spambots \#1} group. The low values of F-Measure and Mathews Correlation Coefficient (MCC), respectively 0.288 and 0.174, are mainly due to the low Recall. In turn, this represents a tendency of labeling \texttt{social spambots \#1} as genuine accounts.

\vskip 1em
\noindent \textbf{Supervised spambot classification.} Among the many supervised classification approaches to spambot detection proposed in recent years by Academia, we decided to experiment with the one presented by C. Yang \textit{et al.} in~\cite{yang2013}, since it focuses on the detection of \textit{evolving} Twitter spambots. 
Thus, it is interesting to evaluate if the system recently presented in~\cite{yang2013} is actually able to detect the sophisticated social spambots. 
This supervised system provides a machine learning classifier that infers whether a Twitter account is genuine or spambot by relying on account's relationships, tweeting timing and level of automation. We have reproduced such a classifier by implementing and computing all the features proposed in~\cite{yang2013}, and by training the classifier with its original dataset. Results in Table~\ref{tab:results} show that the system fails to correctly classify the novel social spambots. Similarly to the results of the BotOrNot? service, the worst results of this system in both \texttt{test set \#1} and \texttt{test set \#2} are related to the Recall metric. This means that also this classifier labeled social spambots as genuine accounts.

\vskip 1em
\noindent \textbf{Unsupervised spambot detection via Twitter stream clustering.} Our initial claim, supported by preliminary work~\cite{ferrara2016, zhang2016}, is that social spambots might be so sophisticatedly designed to make it very difficult to distinguish them from genuine accounts, if observed one by one. If demonstrated, this claim would imply that supervised classification approaches are intrinsically worse than unsupervised ones for the detection of social spambots. For this reason, 
we have also experimented with unsupervised approaches for spambot detection.
The approach in~\cite{miller2014} considers vectors made of 126 features extracted from both accounts and tweets as input of modified versions of the DenStream~\cite{cao2006} and StreamKM++~\cite{ackermann2012} clustering algorithms,  to cluster feature vectors of a set of unlabeled accounts. 
We have implemented the system 
proposed in~\cite{miller2014} 
to cluster the accounts of our 2 test sets.
As shown in Table~\ref{tab:results}, this achieved the worst performances among all those that we have benchmarked in this study. Low values of both Precision and Recall mean incomplete and unreliable spambot detection. Among the 126 features, 95 are based on the textual content of tweets. However, novel social spambots tweet contents similar to that of genuine accounts (e.g., retweets of genuine tweets and famous quotes). For this reason, an approach almost solely based on tweet content will not be able to achieve satisfactory results.

\vskip 1em
\noindent \textbf{Unsupervised spambot detection via graph clustering.} The approach in~\cite{ahmed2013} exploits statistical features related to URLs, hashtags, mentions and retweets. Feature vectors generated in this way are then compared with one another via an Euclidean distance measure. Distances between accounts are organized in an adjacency matrix, which is later used to construct an undirected weighted graph of the accounts. Then, graph clustering and community detection algorithms are applied in order to identify groups of similar accounts. Graph clustering is done by employing the \textit{Markov cluster algorithm} (\textit{MCL})~\cite{van2008}. 
We fully implemented this solution and we experimented with our datasets. However, the approach failed to identify 2 distinct clusters, since accounts of both our test-sets were assigned to a single cluster. We also performed a grid search simulation in order to test the best parameter configuration for \textit{MCL}\footnote{\scriptsize{\textit{MCL} admits 2 fundamental parameters: \textit{inflation} and \textit{expansion}.}}, but to no avail. To achieve effective detection results, instead of the \textit{MCL},
we adopted the \textit{fastgreedy} community detection algorithm~\cite{clauset2004}.  As
reported in Table~\ref{tab:results}, our modified implementation proved
effective in detecting social spambots, with an MCC = 0.886 for
\texttt{test set \#1} and MCC = 0.847 for \texttt{test set \#2}.

%%% End:  

\makeatletter{}
\section{Emerging trends}
\label{sec:newtrends}

\makeatletter{}
\begin{table}[t]
	\scriptsize
    \centering
     \begin{tabularx}{1\columnwidth}{p{0.36\columnwidth}cp{0.26\columnwidth}c}
	\toprule
      	\multicolumn{2}{c}{\textbf{established work}} & \multicolumn{2}{c}{\textbf{emerging trends}}\\
      	\midrule
      		\mbox{Yardi, Romero \textit{et al.}~\cite{yardi2009}}								& 2009	& \mbox{Beutel, Faloutsos} \mbox{\textit{et al.}~\cite{Beutel:2013,jiang2015,Giatsoglou2015,jiang2016}}	& 2013-16 \\		\mbox{Benevenuto \textit{et al.}~\cite{benevenuto2009,benevenuto2010,Ghosh:2012}}	& 2009-12& Cao \textit{et al.}~\cite{Cao:2014}													& 2014 \\
		K. Lee, Caverlee \mbox{\textit{et al.}~\cite{Lee:2010,leeKyumin2011}}				& 2010-11	& Yu \textit{et al.}~\cite{yu2015}														& 2015 \\
		Stringhini \mbox{\textit{et al.}~\cite{Stringhini:2010,Stringhini:2012,Stringhini:2013}}	& 2010-13	& Viswanath, Mislove, Gummadi \textit{et al.}~\cite{viswanath2015}							& 2015 \\
		Viswanath, Mislove, Gummadi \textit{et al.}~\cite{viswanath2011}				& 2011	& Cresci \textit{et al.}~\cite{IntSys2015}												& 2016 \\
		Stein \textit{et al.}~\cite{stein2011}										& 2011	& & \\
		Thomas \textit{et al.}~\cite{ThomasGMPS11}								& 2011	& & \\
		Gao \textit{et al.}~\cite{GaoCLPC12}										& 2012	& & \\
		Cao \textit{et al.}~\cite{cao2012}										& 2012	& & \\
		Xie \textit{et al.}~\cite{xie2012}											& 2012	& & \\
		C. Yang \textit{et al.}~\cite{yang2013}									& 2013	& & \\
		Wang \textit{et al.}~\cite{wang2013}										& 2013	& & \\
		\mbox{S. Lee \textit{et al.}~\cite{Lee:2013,ComCom14}}						& 2013-14	& & \\
		Z. Yang \textit{et al.}~\cite{yang2014uncovering}						        & 2014      & & \\
		Liu \textit{et al.}~\cite{weibo14}											& 2014	& & \\
		Paradise \textit{et al.}~\cite{paradise2014}									& 2014	& & \\
		Cresci \textit{et al.}~\cite{DASec:2014,cresci2015}							& 2014-15	& & \\
		Ferrara \textit{et al.}~\cite{botornot,davis2016}				                                 & 2014-16& & \\
	\bottomrule
	\multicolumn{4}{l}{
	\begin{minipage}[t]{.975\columnwidth}		{\scriptsize This table does not aim to be complete, but rather to testify the emergence of a new trend of research.} 	\end{minipage}}
    \end{tabularx}
\caption{\small Recent work in spambot detection.\label{tab:survey}}   
\end{table}

\begin{figure*}
  \centering
  \subfigure[\label{fig:joindate}]{\includegraphics[width=0.22\textwidth]{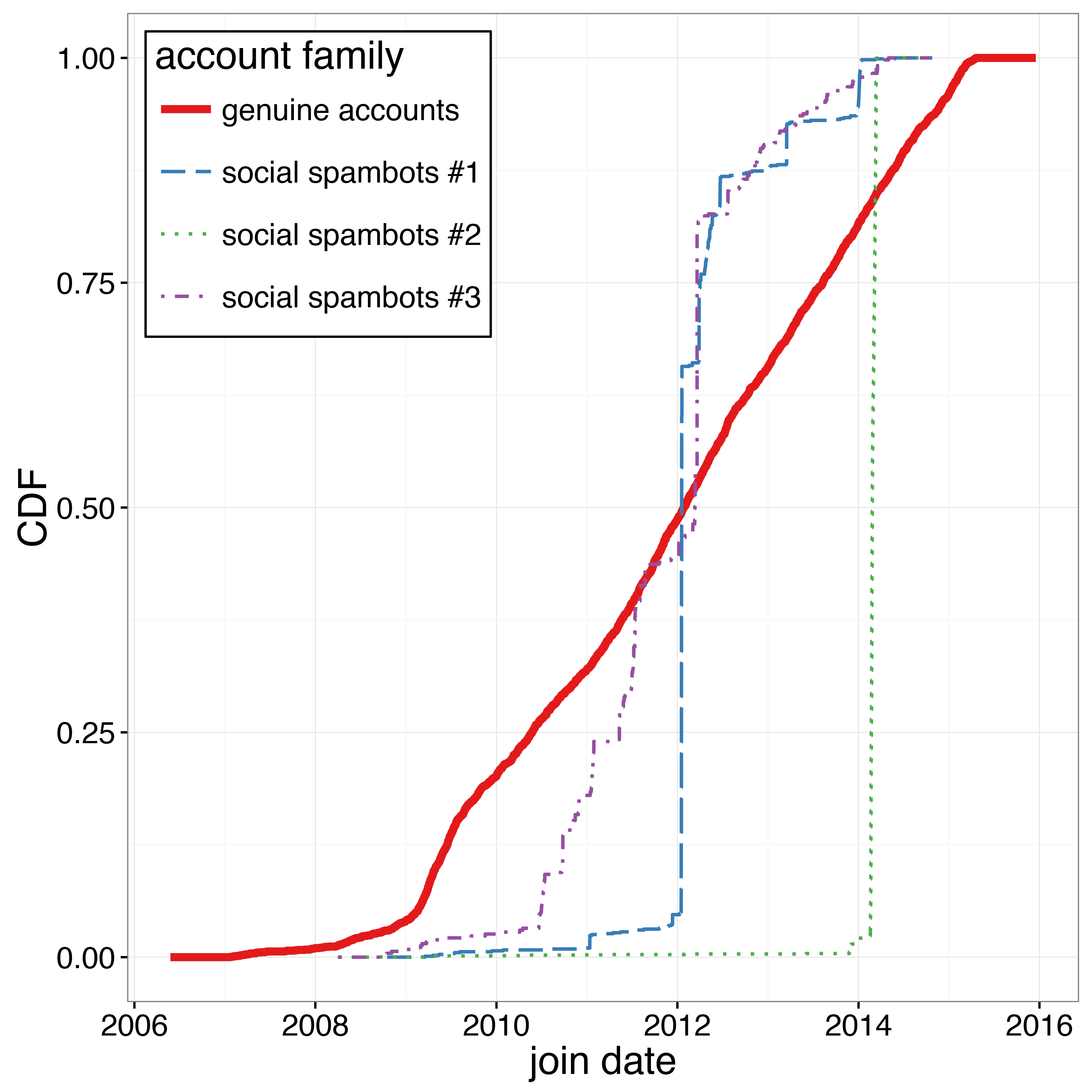}}
  \subfigure[\label{fig:followers}]{\includegraphics[width=0.22\textwidth]{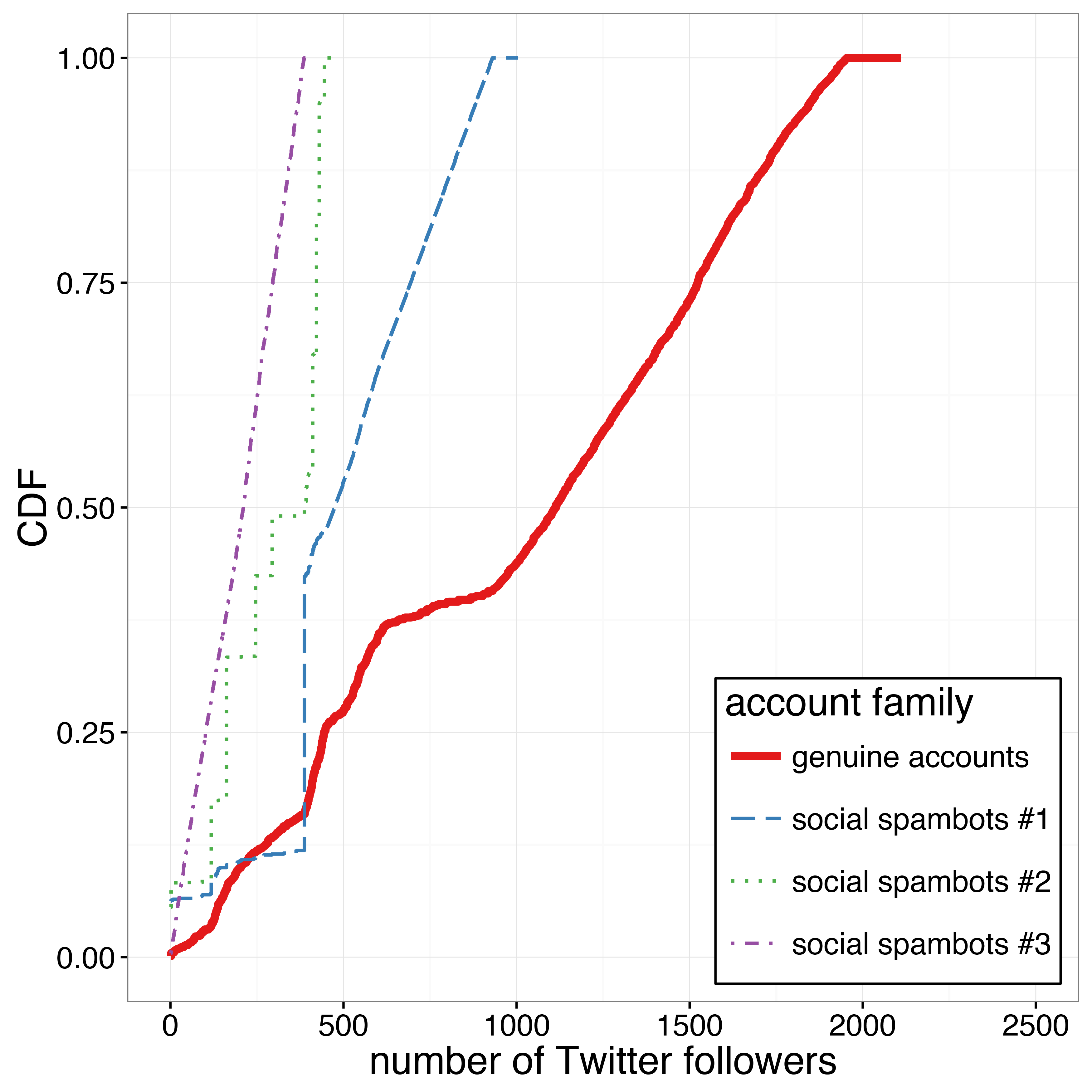}}
  \subfigure[\label{fig:lcs-human-vs-bot}]{\includegraphics[width=0.22\textwidth]{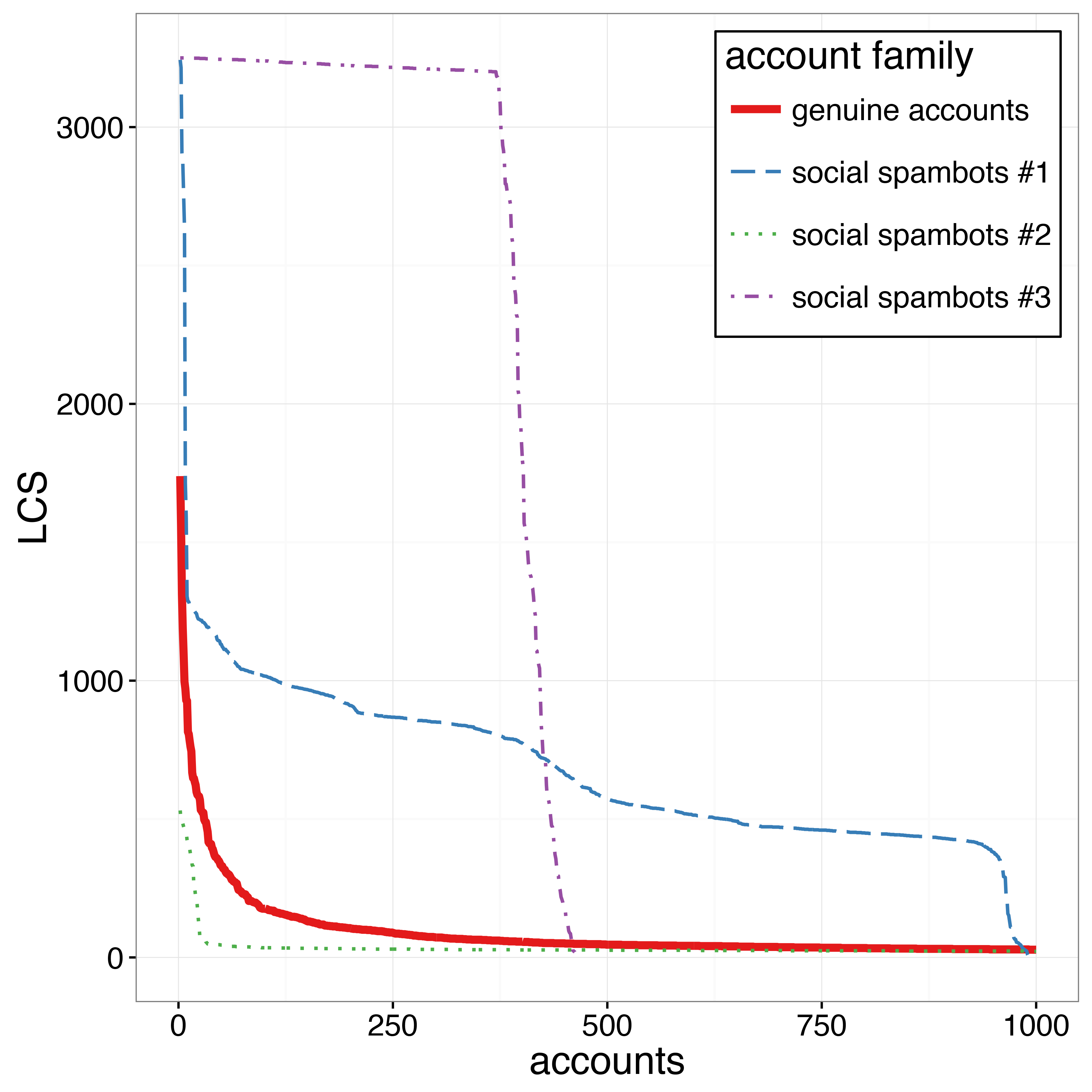}}
  \caption{\small Distribution of join date, number of followers and LCS for genuine and social spambots accounts.
  \label{fig:distribution-join-foll}}
\end{figure*}

As shown in Table~\ref{tab:results}, the established works benchmarked in Section~\ref{sec:theothers} largely failed to detect the new wave of social spambots, one notable exception being the unsupervised approach proposed in~\cite{ahmed2013}. These results call for novel analytic tools able to keep pace with the latest evolutionary step of spambots. Thus, in this section we revise the most recent literature on spambots detection aiming to answer the research question:

\textbf{RQ5 --} \textit{Is it possible to find new dimensions over which to fight and overcome the novel social spambots?}

An answer to this question might be uncovered by reviewing the evolution of research efforts towards the detection of malicious accounts in OSNs.
Such analysis highlights that traditional spambot detection systems typically relied on the application of off-the-shelf machine learning algorithms on the accounts under investigation.
Indeed since 2009, most of the works in this field were focused on designing machine-learning features capable of maximizing detection performances of well-known algorithms, such as SVM, Decision Trees, Random Forests, and more. However, since 2013, a number of research teams independently started to formalize new approaches, also from the algorithmic point of view, for detecting the coordinated and synchronized behavior that characterizes groups of automated malicious accounts~\cite{Beutel:2013}. In Table~\ref{tab:survey} we grouped such techniques and we labeled them as {\it emerging trends}. Despite being based on different key concepts, all these systems propose novel algorithmic solutions and investigate groups of accounts as a whole, marking a significant difference with the previous literature. Table~\ref{tab:emerging} reports on the new concepts introduced by this emerging body of work. 

In order to derive even more insights into the characteristics that might make these emerging solutions successful against current, and possibly future, spambots, in the following we discuss and experiment with the systems proposed in~\cite{viswanath2015, IntSys2015}.

\vskip 1em
\noindent \textbf{Tamper Detection in Crowd Computations.} The contribution by Viswanath \textit{et al.} in~\cite{viswanath2015} checks whether a given group of accounts (e.g., retweeters of another account, reviewers of a venue on {\it Yelp}) contains a subset of malicious accounts.
The intuition behind the methodology is that the statistical distribution of reputation scores (e.g., number of friends and followers) of the accounts participating in a tampered computation
significantly diverge from that of untampered ones. The detection of a tampered computation is performed by computing the Kullback-Leibler distance between the statistical distribution of a given reputation score for the computation under investigation with that of a reference -- untampered -- computation. If such a distance exceeds a given threshold, the computation under investigation is labeled as tampered.
 To test this technique against social spambots, we have computed the statistical distribution of the two reputation scores used in~\cite{viswanath2015} (join date and number of followers) for the genuine and the social spambots accounts of our datasets. The results are in figures~\ref{fig:joindate} and~\ref{fig:followers}. Whereas the genuine accounts feature distributions that almost uniformly span across the possible range of values, social spambots have anomalous distributions. Thus, the technique proposed in~\cite{viswanath2015} is  capable of spotting the differences between groups of genuine accounts and the new wave of social spambots. However, the technique cannot \textit{directly} spot the tampering accounts, and, thus, detection and removal of the single accounts must be performed using a separate methodology.

\makeatletter{}
\begin{table}[t]
	\scriptsize
    \centering\bgroup
\def\arraystretch{1.3}
    \begin{tabular}{p{0.45\columnwidth}p{0.45\columnwidth}}
	\toprule
      	\textbf{work} & \textbf{key concept} \\
      	\midrule
      		Beutel, Faloutsos \textit{et al.}~\cite{Beutel:2013,jiang2015}				& detection of lockstep behaviors \\
		Beutel, Faloutsos \textit{et al.}~\cite{Giatsoglou2015,jiang2016}				& anomalies in synchronicity and normality \\
		Cao \textit{et al.}~\cite{Cao:2014}									& detection of loosely synchronized actions \\
		Yu \textit{et al.}~\cite{yu2015}										& detection of latent group anomalies in graphs \\
		Viswanath, Mislove, Gummadi \textit{et al.}~\cite{viswanath2015}			& distance between distributions of reputation scores \\
		Cresci \textit{et al.}~\cite{IntSys2015}								& similarity between digital DNA sequences \\
	\bottomrule
    \end{tabular}
  \egroup
  \caption{\small Key concepts of emerging trends.\label{tab:emerging}}   
\end{table}

%%% End:  

\vskip 1em
\noindent \textbf{Digital DNA for social spambots detection.} Similarly to~\cite{viswanath2015}, the technique in~\cite{IntSys2015} analyses a group of accounts, for detecting possible spambots among them. Authors introduced a bio-inspired technique to model online users behaviors by so-called ``digital DNA" sequences. Extracting digital DNA for an account means associating that account to a string that encodes its behavioral information. Digital DNA sequences are then compared between one another to find anomalous similarities among sequences of a subgroup of accounts. The similarity among digital DNA sequences is computed in~\cite{IntSys2015} by measuring the Longest Common Substring (LCS) -- that is, the longest DNA substring shared by all the accounts of the group. Accounts that share a suspiciously long DNA substring are then labeled as spambots. Notably, although working at group level, \cite{IntSys2015} is capable of spotting single spambot accounts. For this reason, we have been able to compare this technique with the ones previously benchmarked in Table~\ref{tab:results}. Applying the technique to our datasets, the similarity curve of genuine accounts is significantly different from that of social spambots, as shown in Figure~\ref{fig:lcs-human-vs-bot}. More specifically, as measured by the LCS metric, \texttt{social spambots \#1} and \texttt{\#3} feature a level of similarity much higher than that of genuine accounts. 
Results reported in Table~\ref{tab:results} demonstrate that the digital DNA-based technique~\cite{IntSys2015} achieves excellent detection performances.

\vskip 1em
\noindent \textbf{Spambot detection: the way ahead.} The other systems listed in Table~\ref{tab:emerging},  similarly to~\cite{viswanath2015, IntSys2015}, also focus on those group characteristics that are particularly suitable for discriminating between malicious and genuine accounts, like, e.g., measuring the \textit{synchronicity} and the \textit{normality} of such groups~\cite{Giatsoglou2015,jiang2016}. Focusing on groups has the advantage that, no matter how sophisticated a single spambot can be, a large enough group of spambots will still leave traces of automation, since they do have a common goal (e.g., increasing someone's reputation score). By performing analyses at group level, this emerging trend might be able to significantly raise the bar for social spambots to evade detection. In addition, a further property -- and advantage -- of this recent research wave is that it proposes ad-hoc detection algorithms, rather than adopting generic and off-the-shelf machine learning algorithms.

The compelling features of the emerging techniques listed in this section represents a fertile ground for fighting the novel social spambots. We can observe a paradigm-shift for research and development of spambot detection systems, which may exploit the new concepts to achieve better resilience and robustness and to withstand the next evolution of social media spambots.

%%% End:  

\makeatletter{}
\section{Concluding remarks}
\label{sec:conclusions}
Our long-lasting experiment on malicious accounts survival rate in Twitter demonstrated that spambot detection is still an open issue. 
Moreover, the already difficult problem to detect spambots in  social media is bound to worsen, since the emergence of a new wave of so-called social spambots.
By accurately mimicking the characteristics of genuine users, these spambots are intrinsically harder to detect than those studied by Academia in the past years. In our experiments, neither humans nor state-of-the-art spambot detection applications managed to accurately detect the accounts belonging to this new wave of spambots.
Indeed, our experiments highlighted that the majority of existing automated systems, as well as crowdsourcing, erroneously label social spambots as genuine (human-operated) accounts.
We demonstrated the need for novel analytic tools capable of turning the tide in the arms race against such sophisticated spambots.
One promising research direction stems from the analysis of collective behaviors. We highlighted a few emerging approaches that analyze groups as a whole, rather than individuals.
The promising outcome of these novel approaches clearly  indicates that this is a favorable research avenue.

%%% End:  

\makeatletter{}
\section{Acknowledgements}
\label{sec:ack}
This research is supported in part by the EU H2020 Program under the
schemes \texttt{INFRAIA-1-2014-2015: Research Infrastructures}
grant agreement \#654024 \textit{SoBigData: Social Mining \& Big
Data Ecosystem} and \texttt{MSCA-ITN-2015-ETN} grant agreement \#675320 \textit{European Network of Excellence in Cybersecurity (NECS)}.
Funding has also been received from the Registro.it IIT-CNR project \textit{MIB (My Information Bubble)}, from Fondazione Cassa di Risparmio di Lucca that partially finances the regional project \textit{Reviewland}, and from the MIUR (Ministero dell'Istruzione, dell'Universit\`{a} e della Ricerca) and Regione Toscana (Tuscany, Italy) funding the \textit{SmartNews: Social sensing for Breaking News} project: \texttt{PAR-FAS 2007-2013}.
\begin{figure}[h]
  \centering
{\includegraphics[width=.995\columnwidth]{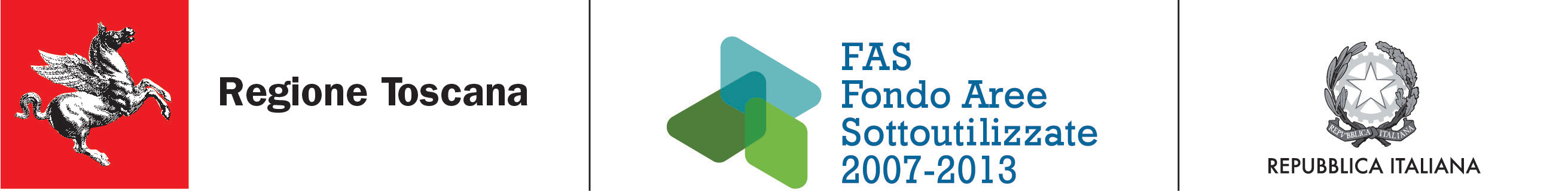}}
\end{figure}

%%% End:  

\balance
\bibliographystyle{abbrv}
\bibliography{references_short} 

\end{document}